\documentclass{aa}
\usepackage{graphicx}

\begin{document}

  \title{
          A Mosaic of the Coma Cluster of Galaxies with XMM-Newton 
  \thanks{
          Based on observations with XMM-Newton, an ESA Science Mission
          with instruments and contributions directly funded by ESA Member
          States and the USA (NASA)}
   }

\author{Ulrich G. Briel,\inst{1} 
        J. Patrick Henry,\inst{1,2} David H. Lumb,\inst{3} 
        Monique Arnaud,\inst{4} Doris Neumann,\inst{4} 
        Nabila Aghanim,\inst{5} Rene Gastaud,\inst{4} 
        Jonathan P. D. Mittaz,\inst{6} Timothy P. Sasseen,\inst{7}
        W. Tom Vestrand\inst{8}
        }

  \offprints{U. G. Briel, \email{ugb@mpe.mpg.de}}

 \institute{$^1$~Max-Planck-Institut f\"ur extraterrestrische Physik,
            D--85748 Garching, Germany\\
            $^2$~Institute for Astronomy,
            2680 Woodlawn Drive, Honolulu, Hawaii 96822,  USA\\
            $^3$~XMM Science Operations Centre, 
            Space Science Division ESTEC, Postbus299, 2200AG
            Noordwijk, Netherland \\
            $^4$~CEA/DSM/DAPNIA Saclay, Service d'Astroph., L'Orme
            de Merisiers Bat 709., 91191 Gif-sur-Yvette, France \\
            $^5$~IAS-CNRS, B\^atiment 121, Universit\'e Paris Sud 91405 Orsay Cedex \\
            $^6$~Mullard Space Science Laboratory, Holmbury St. Mary, Surrey,
  	    RH6 6NT, UK \\
            $^7$~Dept. of Physics, University of California at 
            Santa Barbara, Santa Barbara, CA 93106 \\ 
            $^8$~NIS-2, MS D436, Los Alamos National Laboratory,
            Los Alamos, NM 87545, USA \\ }

   \date{Received October 6, 2000 / Accepted November 4, 2000}

\authorrunning{U.G.Briel et al.}

\titlerunning{A Mosaic of the Coma Cluster of Galaxies with XMM-Newton }

%%%%%%%%%%%%%%%%% BEGINN ABSTRACT %%%%%%%%%%%%%%%%%%%%%%%%%%%%%%%%

\abstract{
The \object{Coma} cluster of galaxies was observed with XMM-Newton in 12 
partially overlapping pointings. We present here the resulting 
X-ray map in different energy bands and discuss the large scale 
structure of this cluster. Many point sources were found throughout 
the observed area, at least 11 of them are coincident with bright 
galaxies. We also give a hardness ratio
map at the so far highest angular resolution obtained for a cluster 
of galaxies. In this map we found soft regions at the position of
bright galaxies, little variation in the central 15 arcmin, but some
harder regions north of the line \object{NGC 4874} -- \object{NGC 4889}.
}

%%%%%%%%%%%%%%%%% END ABSTRACT %%%%%%%%%%%%%%%%%%%%%%%%%%%%%%%%

   \maketitle

%%%%%%%%%%%%% BEGINN INTRODUCTION %%%%%%%%%%%%%%%%%

\section{Introduction}

The \object{Coma} cluster of galaxies was one of the performance verification 
targets of XMM-Newton. The aim was to verify XMM-Newton's ability to map large 
extended X-ray sources. This is not an easy task because errors can occur in
several areas. 

The telescope vignetting plays a major role when overlapping
the same region of an extended source, measured at different off-axis angles.
Because of the vignetting, the surface brightness of a region measured off 
axis seems to be lower than the same region measured on axis. The reason for
that is the increasing loss of mirror collecting area with increasing 
off-axis angle, which in addition is energy dependent. By de-vignetting the
different observations and normalizing the observations to the obser\-ving times
the surface brightnesses of the same source regions are corrected and must show 
the same number of counts/arcmin$^2$/sec. If done correctly,
detector-edges or regions of detector gaps should not be visible as 
inhomogenities in the surface brightness of the merged observations.

\begin{table*}[t]
\caption[ ]{Journal of Observations}
%\begin{flushleft}
\begin{tabular}{llccccc}
\hline
\rule{0mm}{4mm}
%\\[0.5ex]
     & Name of     &          &           &  \multicolumn{3}{c}{pn-Camera observing times (ksec)}\\
\multicolumn{1}{c}{Date} & Observation & RA(2000) & DEC(2000) & planned     & 
\multicolumn{1}{c}{performed} & \multicolumn{1}{c}{effective} \\
\hline
\rule{0mm}{4mm}
2000 May 29 \    & Coma center & 12 59 46.7 & 27 57 00 & 15.0 & 22.7 & 15.4 \\
2000 June 21/22  & Coma 1      & 12 56 47.7 & 27 24 07 & 25.0 & 29.2 & 24.0 \\ 
2000 June 11/12  & Coma 2      & 12 57 42.5 & 27 43 38 & 25.0 & 34.8 & 34.7 \\
2000 June 27/28  & Coma 3      & 12 58 32.2 & 27 24 12 & 25.0 & 27.0 & 14.3 \\
2000 June 23     & Coma 4      & 13 00 04.6 & 27 31 24 & 25.0 & 20.0 & (11.4) \\
2000 May 29      & Coma 5      & 12 59 27.5 & 27 46 53 & 20.0 & 25.0 & 10.0 \\ 
2000 June 12     & Coma 6      & 12 58 50.0 & 27 58 52 & 20.0 & 14.8 &  3.7 \\
2000 June 24     & Coma 7      & 12 57 27.7 & 28 08 41 & 25.0 &  8.1 & (3.4) \\
not observed yet & Coma 8      & 13 01 25.6 & 27 43 53 & 26.0 &   -- &   -- \\
2000 June 12     & Coma 9      & 13 00 32.7 & 27 56 59 & 20.0 & 26.3 & 21.3 \\
2000 June 22     & Coma 10     & 12 59 38.4 & 28 07 40 & 20.0 & 26.3 & 16.6 \\
2000 June 24     & Coma 11     & 12 58 36.5 & 28 23 56 & 25.0 & 32.1 & 15.9 \\
not observed yet & Coma 12     & 13 01 50.2 & 28 09 28 & 25.0 &   -- &   -- \\
not observed yet & Coma 13     & 13 00 36.5 & 28 25 15 & 25.0 &   -- &   -- \\
2000 June 22     & Coma bkgd   & 13 01 37.0 & 27 19 52 & 30.0 & 22.7 & 20.1 \\
\hline
\end{tabular}
%\end{flushleft}
\end{table*}

The backgrounds in the image are problematic.  The part of the background
that is produced by particles inside the detector is not distinguishable 
from X-rays, is not vignetted, and is in general distributed 
homogeneously over the detector face (except for the lower end of the 
spectral range and around the detector intrinsic emission lines around 8 keV,
see Briel et al. \cite{ub00}). 
Hence this component must be subtracted
before de-vignetting is applied. 
The extragalactic X-ray background is vignetted in the same manner as 
diffuse emission from an extended source and may therefore be de-vignetted in 
the same manner.
But there is a third component in the background, the low energy protons,
that are 
again indistinguishable from X-rays. They come through the mirror system
and show a different vignetting compared to X-rays (see Str\"uder et al. 
\cite{ls01}). In addition, their spectrum and their flux is time variable. 
Because of the different vignetting, it is essential to subtract a ``proton
image" from the measured image before de-vignetting is applied, especially
when the surface brightness of the proton induced background is in the 
same order of magnitude as the surface brightness of the extended source.
  
\object{Coma} was chosen because it is a well studied source at all wavelength 
ranges. With a diameter of the main cluster, which is several times larger 
than the FOV of the EPIC cameras of about 26 arcmin, it is
necessary to mosaic the cluster with XMM-Newton if the large scale structure
and dynamics of the cluster are to be investigated. Although \object{Coma} 
was long believed to be the archetype of a relaxed cluster of galaxies, the 
X-ray image taken during the Rosat All Sky Survey revealed the complex large 
scale structure, especially the subgroup around \object{NGC 4839} probably 
falling into the main cluster (Briel et al. \cite{ub92}). 
Subsequent pointed observations with Rosat showed that \object{Coma} has 
irregular structure on different scales and was most
probably formed by hierarchical clustering of several subunits, which are
still visible in their X-ray emission (White et al. \cite{sw93}). 
Observations with ASCA, and also the pointed Rosat observation, showed evidence
of temperature structure in \object{Coma} (Honda et al. \cite{hh96}; 
Briel \& Henry \cite{ub98}; Donnelly et al. \cite{rd99}; 
Watanabe et al. \cite{mw99}).
The interpretation of the temperature structure favors the ongoing merging  of 
the subgroup around \object{NGC 4839} with the main cluster, a scenario which 
is still debated. 

In this letter we will present the result of merging 12 partially overlapping
observations of the \object{Coma} cluster with the EPIC pn-CCD-detector on board
XMM-Newton. In Section 2 we will describe the observations and present
the data reduction and the method of merging the individual observations.
In Section 3 we will discuss some of the many point sources in the image,
the large scale structure of the X-ray surface brightness distribution, and 
the hardness ratio map of the inner cluster region.

%%%%%%%%%%%%% END INTRODUCTION %%%%%%%%%%%%%%%%%

%%%%%%%%%%%%% BEGIN OBSERVATIONS AND DATA REDUCTION %%%%%%%%%%%%%%%%%

\section{Observations and Data Reduction}

The region of the \object{Coma} cluster of galaxies was observed by the 
XMM-Newton Satellite (Jansen et al. \cite{fj01}) during the performance 
verification phase. In Table 1 we show the journal of the observations with 
information on the pointing directions and observing times.

The two MOS-CCD cameras   (Turner et al. \cite{mt01}) were operated in their
full-frame mode, which  we believe to be well-calibrated for spectral analysis.
However due to operational difficulties, less data coverage of the 
\object{Coma} region was obtained in these cameras. The pn-CCD camera 
(Str\"uder et al. \cite{ls01}) was operated in the extended full frame mode 
(see the above papers for a description of the different modes). 
This latter mode provides potentially the best imaging
performance of the pn-CCD camera, but at this date its spectral
calibration is less reliable than that of the MOS observation. The
analysis of this paper therefore contrasts with that of Arnaud et al.
(\cite{ma01}), in that we concentrate more on the gross imaging properties 
and less on the spectral imaging.  The hardness ratio map that we present,
based on flux estimate in wide energy bands, is not sensitive to the
remaining PN calibration uncertainties. For all observations, the medium 
filter was used (see Turner et al. \cite{mt01}).

\begin{figure*}[t]
\centering
\includegraphics[width=14.0truecm,angle=0,clip]{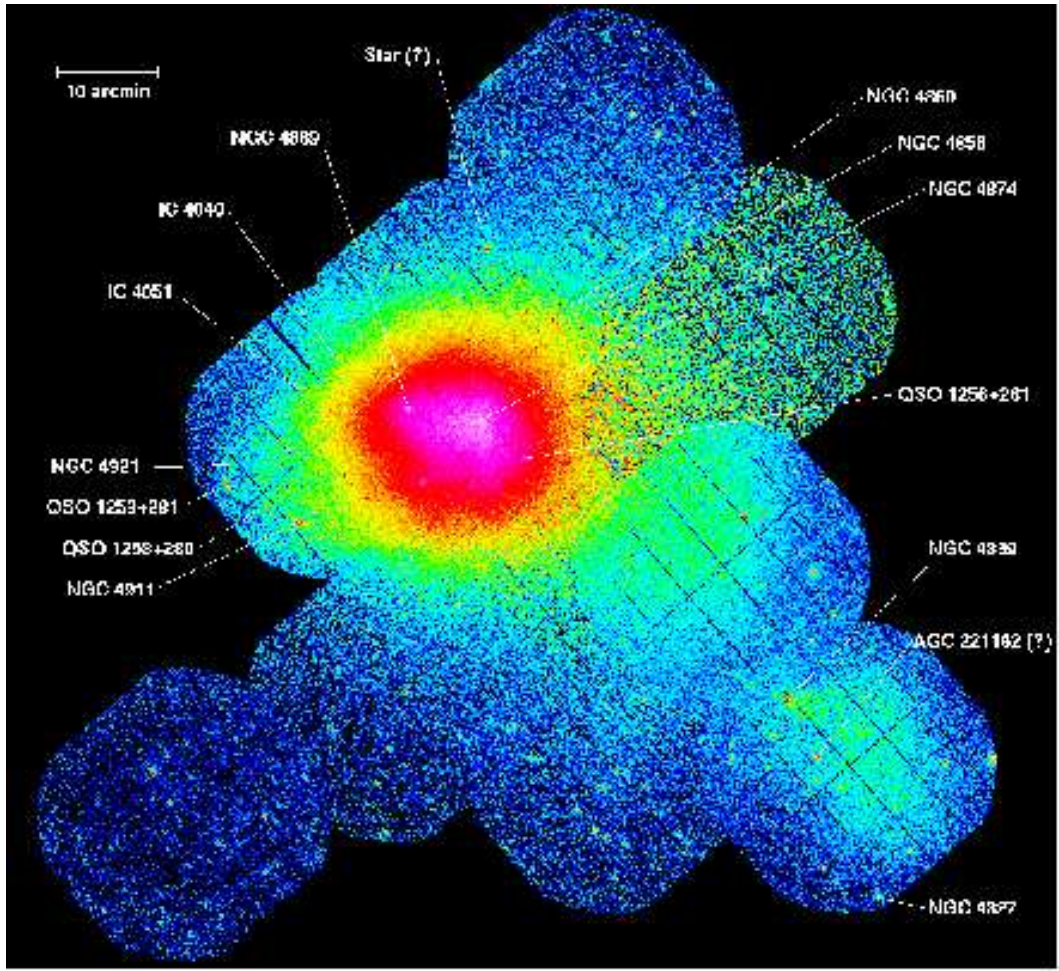} 
\caption[ ]{
The merged EPIC-pn image of the \object{Coma} cluster of galaxies in the
0.3 to 2.0 keV energy band. Indicated are the tentative identifications 
of point sources.}
\end{figure*}

\begin{figure*}[t]
\centering
\includegraphics[width=14.0truecm,angle=0,clip]{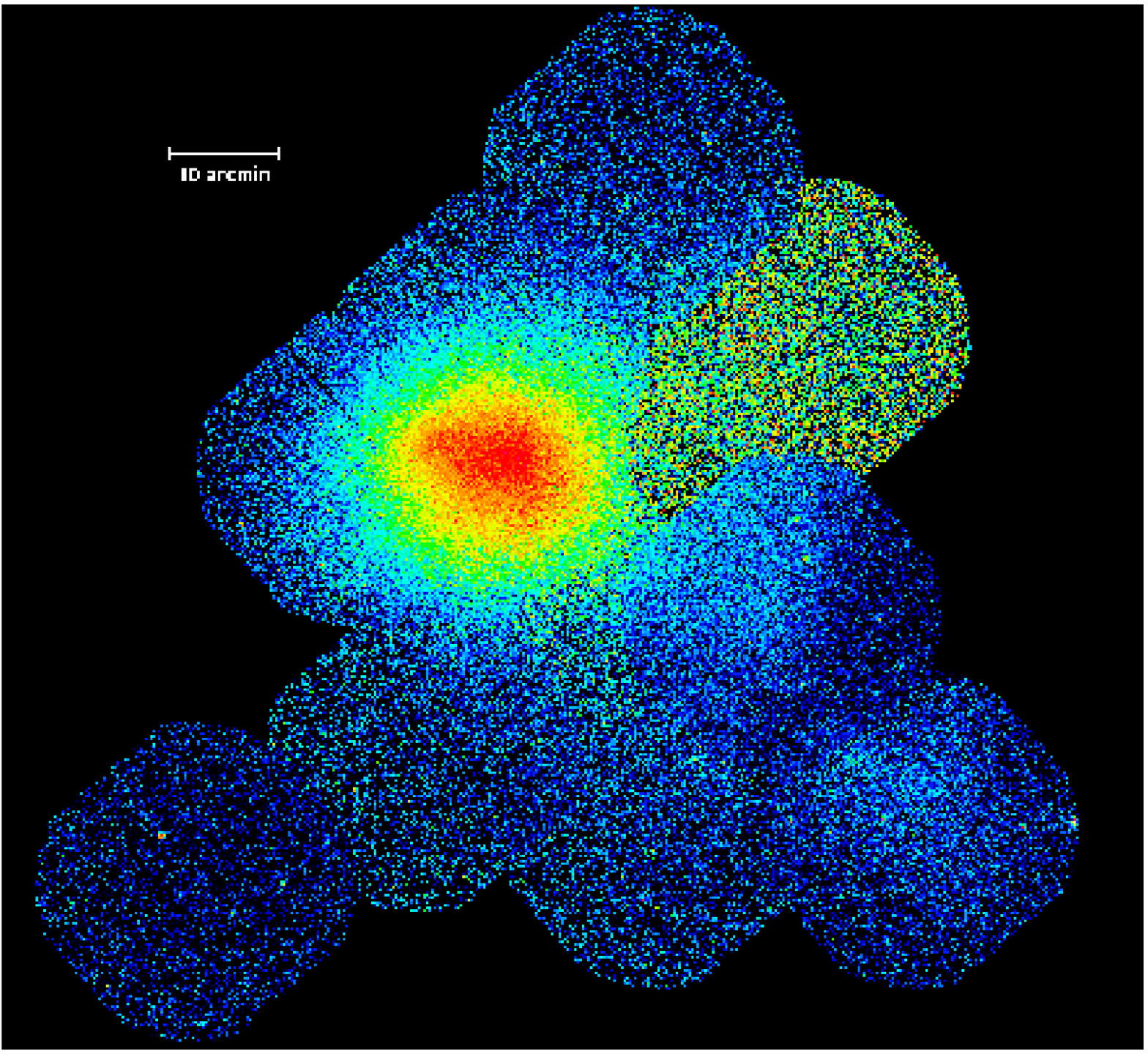}
\caption[ ]{
The merged EPIC-pn image of the \object{Coma} cluster of galaxies in the 
2.0 to 5.0 keV energy band.}
\end{figure*}

During the observations the proton induced background often varied, and 
in flares reached count rates up to several hundred counts/sec in the energy 
band from 0.15 to 10 keV. To increase the signal-to-noise ratio and also
to minimize the above described problems of the de-vignetting of residual
proton induced events, we binned the 10-15 keV events in 100 sec time bins.
From this light curve, times of bins with count rates above 100 counts/100 
sec were identified and excluded from further processing. We have chosen
the high energy region because we do not expect many events from \object{Coma} 
in that band. This rate criterion gave in all but Coma 4, 6, and 7 
reasonably long effective exposure times, which are given in Table 1.
The rate criterion for Coma 4 and 7 were 200 counts and 800 counts 
respectively. We did include these high background times in order
to get the mosaic of \object{Coma} as complete as possible. We do have to be 
careful with the interpretation of those two fields. Since the Coma 6 
field covers part of the center of the cluster, we did apply the 
stringent rate criterion, therefore 3.7 ksec exposure time is
usable, which is only 25\% of the total time pointed at this position.

\begin{figure*}[t]
\centering
\includegraphics[width=12.0truecm,angle=0,clip]{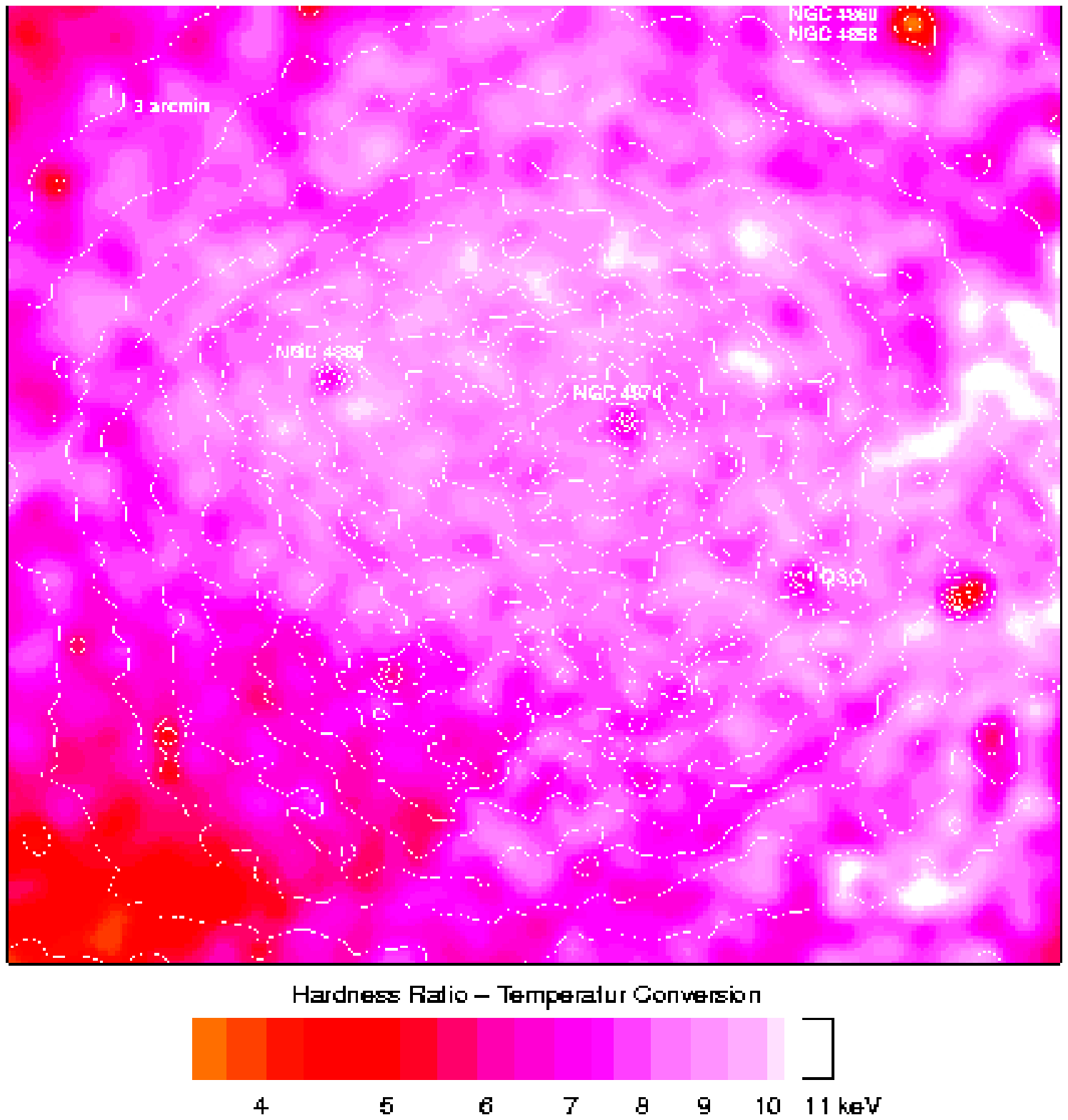}
\caption[ ]{
The hardness ratio map of the inner \object{Coma} region in the 
energy bands from 0.3-2 keV and 2-7.15 keV. The overlain 
contours are from the smoothed surface brightness distribution
in the energy band from 0.3-2 keV
}
\end{figure*}

In addition to the above screening, we also excluded ``warm" pixels from the 
images by checking each indivi\-du\-al observation in detector coordinates
and searching for single pixels with statistically significant higher
numbers of events compared to adjacent pixels. 

Images were produced in three different energy bands: 0.3 - 2 keV, 2 -
5 keV and 5 - 7.15 keV. We stopped at 7.15 keV in order to avoid the
strong detector intrinsic emission lines of Ni-K, Cu-K and Zn-K,
spread over approximately 2 keV. At a later time, when we learn more
about the background, the higher energies will be included. An other
handicap was the fact that, at the time of analysing the pn-data the
correction files for the charge transfer inefficiency (CTI) were only
available for the full frame mode (not the extended full frame mode
used for our observations), so they were used to correct for the
CTI. This effects the corrected amplitude of the events at the level
of 5-25 eV depending on the energy and the position in the detector.
But for producing images in the above bands, this systematic error is
negligible. All valid events were used in the images. Valid events are
all single events plus recombined split events, whose position is
defined as the position of the pixel containing the maximum amplitude
(see Dennerl et. al. \cite{kd99} for a detailed explanation). To minimize
the gaps between CCDs, events at the border pixels are also included,
although their amplitude could be slightly under estimated because of
a possibly not recorded split companion. Again, for making images in
wide energy bands, this systematic error is negligible.

In the same energy bands, and with the same selection criteria,
background images were produces from a dedicated ``background event
file". This file contains 132.8 ksec of pn-data accumulated from 5
other observations, like the Lockman Hole (Hasinger et al. \cite{gh01}),
containing no extended X-ray sources and only weak point sources.
The single observations are between 13 and 33 ksec long, spread over 
the 3 months CAL/PV phase of XMM-Newton.  
These data were also screened for flares and warm pixels, and
obvious point sources were taken out. Comparing the single observations 
of the screened event files, the background rate in the energy band from 
0.15 to 10 keV is very stable, the systematic difference between the
minimum and maximum rate is 8\% of the mean. From each \object{Coma} image the
corresponding background image, normalized using the effective
observing times, was subtracted.  Even though we restrict ourselves to
observing times with "low" background, at a distance of 30 arcmin from
the center of \object{Coma} the background is compareable to the surface
brightness of \object{Coma}.  At the center of \object{Coma} the background 
is $\sim$ 1.5\% and 2.5\% of \object{Coma}'s surface brightness in the bands 
0.3 to 2 keV and 2 to 7.15 keV respectively.

In a further step, exposure maps were produced for the different energy
bands, also normalized to the corresponding effective observing times.
These exposure maps contain the energy dependent vignetting. 

The individual images in the lowest energy band were searched for point
sources, and the positions of point sources in overlapping regions 
were then used to align and merge both the background-free X-ray 
images and the exposure maps. In a last step, the merged background 
free X-ray images were divided by the corresponding merged exposure maps.
The result is the surface brightness distribution of the \object{Coma} field
in the different energy bands. In the Figures 1 and 2  we show the 
0.3-2.0 keV image and the 2-5 keV image of \object{Coma} taken with the 
pn camera. 

To visualize better the low surface brightness of the cluster we
smoothed the images with two-dimensional Gaussian filters of variable
sizes (with a sigma from 9 arcsec to 96 arcsec).  We used these
images to look for spectral variations in the central \object{Coma} region. 
A hardness ratio map was produced by calculating the image HR =
(image(2-7.15keV)-image(0.3-2keV))/(image(2-7.15keV)+image(0.3-2keV)).
In Fig. 3 we show the inner 25.5$\times$23.3 arcmin$^2$ of this
hardness ratio map. The overlain contours are from the smoothed surface 
brightness distribution in the 0.3-2 keV band. We performed simulations
with XSPEC, using a Raymond \& Smith model (with the fixed parameters
z = 0.0232, abundance = 0.25 and $N_H$ = 8.95 $10^{19}$ cm$^{-2}$ from
Dickey \& Lockmann \cite{jd90}) and taking into account 
our current understanding of the errors in the preliminary
calibration of the extended full frame mode.  With these simulations we 
converted the hardness ratios into ''temperatures". The resulting conversion 
is indicated as color bar with the corresponding temperatures in Fig. 3.

The dynamic range of the hardness ratio in the map is of the order of 
20\%, corresponding to a ''temperature" variation of $\sim$ 5 keV.
A systematic error of the background model of 10\% would be neglectible
in the inner 15 arcmin diameter. At larger radii the ratio would be 
different by $\sim$ 3\%, corresponding to a ''temperature" difference of 
about 1 keV.

%******************************************************************************

\section{Discussion}

The \object{Coma} cluster is the nearest very rich cluster of galaxies. It is
probably the best studied cluster at all wavelengths and is one of the
brightest extragalactic X-ray sources in the sky. Until now, the
deepest X-ray observation of \object{Coma} is that of White et al. 
(\cite{sw93}) consisting of 4 exposures of about 19 ksec each by the ROSAT
PSPC.  The EPIC pn observations reported here are of comparable
length, thus are $\sim$8 times deeper, while simultaneously extending
over $\sim$3 times wider energy range and with $\sim$4 times better
angular resolution.

Perhaps not surprisingly given the better spatial re\-solution and
deeper observation, the most noticable difference between our XMM
images in Figures 1 and 2 compared to Figure 2 of White et al. (\cite{sw93}) 
is the myriad of newly detected point-like sources. Dow
\& White (\cite{kd95}) discuss the $\sim$25 point sources in the ROSAT
observations of \object{Coma}. There are at least 75 point sources in the
smaller solid angle we have analyzed. We have made a first attempt to
identify these sources by overlaying our X-ray map on an optical image
of the region and then consulting NED when there appears to be a
coincidence. Many bright galaxies seem to be X-ray sources including:
\object{NGC 4827}, \object{4839}, \object{4858}, \object{4860}, 
\object{4874}, \object{4889}, \object{4911}, and \object{4921}, 
and \object{IC 4040} and \object{4051}. 
However some faint galaxies, with magnitudes from 18 to 20, could also 
be identifications based on positional coincidence. At least three quasars, 
\object{QSO 1256+281}, \object{QSO 1258+280},  and \object{QSO 1259+281}, 
are detected. However, the bulk of the sources are unidentified in our 
initial analysis.

The \object{NGC 4911} and \object{NGC 4839} groups contain concentrations 
of point sources at our sensitivity. Dow \& White (\cite{kd95}) and Neumann 
et al. (\cite{dn01}) discuss the particularly interesting morphology of the 
\object{NGC 4839} group.

The large-scale diffuse emission in the soft band shown in Figure 1 is
very similar to the ROSAT observations of White et al. (\cite{sw93}) 
in the same band.  
The X-ray emission of the \object{Coma} cluster is lumpy near its
center but becomes smoother at larger radii. An additional lump,
apparent in the ROSAT observation but not discussed by White et al.
(\cite{sw93}), is directly NE of \object{NGC 4839}. It may be yet another 
group on its way into the main cluster to the ENE. This lump may be preceeding
the \object{NGC 4839} group from the direction of \object{A1367} since 
calculations of cluster growth indicate that the groups that merge to form 
clusters flow along filaments connecting the clusters. The large-scale diffuse
hard emission is very similar to the soft, as Figure 2 shows.

Because the calibration of the spectral response of the extended full frame 
mode of the pn-CCD is not yet available, we do not yet want to perform 
spatially resolved spectral fits. We can, however, make a hardness ratio map 
using wide bands, which we show in Figure 3. While these maps sacrifice 
spectral details, such as temperatures and abundances, they are the highest 
angular ($\sim$40 arcsec) and spatial ($\sim$26 kpc for H$_{\mbox{o}}$
= 50 km s$^{-1}$ Mpc$^{-1}$) resolution spectral maps ever obtained for a 
cluster of galaxies. 

Using the conversion from hardness ratio to temperature, given in Fig. 3, 
we have checked our resulting mean temperature in some regions which are 
known from other measurements. For the over-all temperature within a radius 
of 10 arcmin, centered on \object{NGC 4874}, we determined a temperature of
8.8 keV. The GINGA value is 8.21 $\pm$ 0.09 keV (Hughes et al. \cite{jh93}),
the ASCA value is 9. $\pm$ 0.6 keV (Donnely et al \cite{rd99}) and the 
XMM-MOS value is 8.25 $\pm$ 0.10 keV (Arnaud et al \cite{ma01}). 
We agree to better than 10\% with these previously determined values, 
which would be expected given our understanding of the uncertainties 
of the calibration and of the background model. 
In a further check, we determined the temperature
around \object{NGC 4874} within a radius of 0.5 arcmin and within a ring
from 1 to 5 arcmin. Our result is 7 and 9.2 keV, the MOS result
(Arnaud et al \cite{ma01}, Fig. 8) is 6.6 and 8.6 keV respectively, again
in agreement within 10\%. We found the same kind of agreement on other 
places, taking into account the large area of 3.5 $\times$ 3.5 arcmin$^2$,
used for the MOS data by Arnaud et al. (\cite{ma01}).

The large-scale description of the data in Figure 3 
agrees with Arnaud et al. (\cite{ma01}): there is cool gas to
the south-east and hotter gas to the south-west and west. It agrees
to some extent with Donnely et al. (\cite{rd99}), Watanabe et al. 
(\cite{mw99}) and Briel \& Henry (\cite{ub98}) investigations.
In particular the south-east cool region  was already identified by 
Donnely et al. (\cite{rd99}) and a hot region in the south-west/west
direction, but somewhat further away, was found by Watanabe et al.
(\cite{mw99}).

There are spectral variations on the smallest scales that we can
measure. Softer emission is associated with many compact sources some
of which are associated with galaxies as discussed previously. 
Although the smoothing of the original surface brightness distribution 
smears out the point sources to some extend, a significant drop in the 
radial temperature  profile within 1 arcmin of \object{NGC 4874} was also 
seen by Arnaud et al. (\cite{ma01}), Fig. 8. The Rosat temperature map of
Briel \& Henry (\cite{ub98}) also shows cooler regions around the two
central galaxies (although other cool regions in the center of their
map are not confirmed).  If these soft regions, associated with
galaxies, are extended, we can postulate three possible
explanations. First, they are local potential wells centered on
cluster galaxies as is was discussed by Vikhlinin et al. (\cite{av94}). 
The shallower potential wells could only trap lower temperature gas.
Second, they are the interstellar emission that the cluster galaxy has
been able to retain. Third, integrated emission from low mass X-ray binaries 
which Irwin \& Sarazin (\cite{ji98}) have discussed as the source of 
soft emission from elliptical galaxies. 

Perhaps the most interesting small scale spectral structures are the isolated 
hard spots that do not appear to be associated with any obvious point-like 
X-ray emission.  For the most part they are not associated with an optical 
object in NED. The five hard spots north of the \object{NGC 4889} -- 
\object{NGC 4874} line may be the cause of the hot region in the same general 
location detected by Donnelly et al. (\cite{rd99})  in their ASCA observations. 
It seems that these hot spots are washed out in the temperature map of 
Arnaud et al. (\cite{ma01}) because of the larger (3.5$\times$3.5 armin$^2$) 
integration boxes. 
They might have been seen however in the even  larger integration
region of ASCA, if they are associated with variable sources like QSO's.
But we can not entirely eliminate an instrumental cause  for some of them 
because some are aligned parallel to the CCD orientation and on the boarder 
of the chips.

\begin{acknowledgements}
The XMM-Newton project is supported by the Bundesministerium f\"ur Bildung
und Forschung/Deutsches Zentrum f\"ur Luft- und Raumfahrt (BMFT/DLR),
the Max-Planck Society and the Heidenhain-Stiftung, and also by PPARC,
CEA, CNES, and ASI. \\
J. P. Henry thanks Prof. J. Tr\"umper for the hospitality at the MPE.
We thank the anonymous referee for useful comments, which improved
the paper.
\end{acknowledgements}

\end{document}